\documentclass{iau} 
\usepackage{graphicx}

\title[Astrostatistics and astroinformatics] 
{The changing landscape of astrostatistics and astroinformatics}

\author[Eric D. Feigelson]   
{Eric D. Feigelson} 

\affiliation{Center for Astrostatistics and Department of Astronomy and Astrophysics, Pennsylvania State University, University Park PA 16802 USA \\
email: {\tt edf@astro.psu.edu} }

\pubyear{2017}
\volume{325}  
\jname{Astroinformatics}
\editors{M. Brescia et al., eds.}

\begin{document}

\maketitle

\begin{abstract}
The history and current status of the cross-disciplinary fields of astrostatistics and astroinformatics are reviewed.  Astronomers need a wide range of statistical methods for both data reduction and science analysis.  With the proliferation of high-throughput telescopes, efficient large scale computational methods are also becoming essential.   However, astronomers receive only weak training in these fields during their formal education.  Interest in the fields is rapidly growing with conferences organized by scholarly societies, textbooks and tutorial workshops, and research studies pushing the frontiers of methodology.  R, the premier language of statistical computing, can provide an important software environment for the incorporation of advanced statistical and computational methodology into the astronomical community.  

\end{abstract}


\begin{quote}
\begin{center}
{\bf An aphorism}
{\it 

.~\\
The scientist collects data in order to 

understand natural phenomena

 . \\

The statistician helps the scientist 

acquire understanding from the data 

. \\

The computer scientist helps the scientist

perform the needed calculations (*)

. \\

The individual proficient at all these tasks 

is a data scientist

~ \\
}
\end{center}

{\it (*) help needed only for Big Data }

\end{quote}

\section{The role of statistics and computation in astronomical research}

Astronomers combine telescopic observations of cosmic populations in the effort to understand astrophysical conditions and processes throughout the Universe. Telescopes at all wavebands of light $-$ and recently telescopes for neutrinos and gravitational waves $-$ are pointed at a myriad targets to characterize properties of planets, stars, the Milky Way Galaxy, other galaxies, material between the stars and galaxies, and the Universe as a whole.  In an increasing proportion of studies, a dataset of considerable size is collected.  This might be zero-dimensional photometry, one-dimensional spectra or time series, two-dimensional images, three-dimensional hyperspectral or video images.  The targets may be a single cosmic target, a small sample of targets sharing common properties, or a large heterogeneous sample of targets.   This last class emerges from sensitive wide-field astronomical surveys that have growing importance at all wavebands of astronomy.  

Once the telescope observations are complete, the astronomer faces the task of data analysis.  According to R.\ A.\ Fisher (1922), the brilliant founder of much of 20th century statistics, this is the task of statistics:  
\begin{quote}
In order to arrive at a distinct formulation of statistical problems, it is necessary to define the task which the statistician sets himself: briefly, and in its more concrete form, the object of statistical methods is the reduction of data.  A quantity of data, which usually by its mere bulk is incapable of entering the mind, is to be replaced by relatively few quantities which shall adequately represent the whole, or which, in other words, shall contain as much as possible, ideally the whole, of the relevant information contained in the original data.
\end{quote}

For the early stages of data analysis, I suggest, the astronomer is quite proficient.  The CCD image must be flat-fielded with bias removed and photometrically calibrated to standard stars.  The spectrum must be extracted after removal of cosmic ray streaks,  and drizzled onto a fixed wavelength grid after fitting a polynomial derived from calibration observations of atomic line standards.  The interferometric visibilities must be Fourier transformed into an image or datacube with computationally intensive restoration procedures to treat incomplete coverage in the Fourier plane.  Ancillary information from the instrument is collected and used to improve calibration, point spread functions, and registration to a fixed grid on the sky.  

The astronomer must then engage in Fisher's data reduction, transforming terabytes or petabytes of data with kilobytes of digestible information in the form of tables and figures for communication to the wider scientific community.  This data analysis and reduction is then followed by the more intellectually challenging stage of science analysis.  This can start with 
{\it prima facie} interpretation of the results, but often proceeds with comparing the findings to mathematical models.  These might be simple heuristic models, such as a power law relationship between two variables, or more complex nonlinear and multivariate models derived from astrophysics.  

Astronomy is unusual in the intensity of this last step.  Many fields that collect and interpret data $-$ social sciences such as economics, biological sciences such as ecology or genomics, Earth sciences such as meteorology or seismology $-$ do not have physical models equivalent to elliptical orbits based on Newtonian mechanics or spectral lines based on atomic physics.  Nonlinear regression thus plays an important role in linking astronomical data to astrophysical models and, if the astrophysical theory or past observational studies constrain parameters in advance of a particular study, then Bayesian modeling to estimate posteriors that update priors is used.  In the past few years, astronomers have become among the most active scientific practitioners of Bayesian inference.  

The astronomer and astrophysicist needs a very wide suite of statistical methods.  In observational cosmology, for example, Table 1 shows an association between statistical topics and astronomical problems.  Once a statistical analysis is performed, the scientist should ask questions such as: How reliable are my results (often using P$<$0.003, the 3-sigma criterion)?  Have I used the most effective and reliable statistical procedures for achieving my results?  Do my interpretations depend on heuristic or uncertain models?  Do my conclusions depend on my chosen analysis path?   

\bigskip
\begin{center}
{\bf Statistical Fields for Research in Cosmology}
\begin{tabular}{| ll |} \hline
{\bf Cosmology} & {\bf Statistics} \\
Galaxy clustering		&      			Spatial point processes, clustering \\
Galaxy morphology		  &    			Regression, mixture models \\
Galaxy luminosity function	   &  			Gamma distribution \\
Power law relationships	&      		Pareto distribution  \\
Weak lensing morphology	   &   		Geostatistics, density estimation \\ 
Strong lensing morphology	&      		Shape statistics  \\
Strong lensing timing		     &		Time series with lag \\
Faint source detection		&      		False Discovery Rate \\
SN Ia \& quasar lightcurves    	&	       Nonstationary time series \\
Cosmic microwave background	 ~~~~~	     & 		Markov fields, ICA, etc. \\
$\Lambda$CDM parameters	      	      &		Bayesian inference \& model selection \\
Comparing data \& simulation	&       Uncertainty quantification \\ \hline
\end{tabular}
\end{center}
\bigskip

The role of computation is easily discerned.  The telescopic data are typically passed through a complex instrument-specific software pipeline, and then are subject to flexible analysis for the specific purposes of the research project.  Virtually all astronomers are competent programmers in general purpose languages like Python, C or IDL as well as  discipline-specific software environments developed over many years.  For many problems, analysis is performed on desktop computers with a single CPU and sub-terabyte local storage.  

But increasingly, the dataset has high volume, arrives with high velocity, and exhibits nontrivial variety of structure .... the Three V's of Big Data. But even with moderate-sized datasets, the computational burden may be heavy due to, for example, fitting high-dimensional models, bootstrap-type simulations for model evaluation, or Markov chain Monte Carlo procedures for Bayesian model computations.  For both Big Data and computationally intensive situations, the desktop hardware is insufficient and the astronomer must move into a parallel computation on multi-processor computers, sometimes equipped with high speed Graphical Processing Units.  These supercomputers might reside at the local institution, at the observatory responsible for the telescope, or on the commercial cloud.  

\section{The education gap}

Despite these strong needs, astronomers are ill-prepared for community-wide use of advanced statistical and computational methods.  The research literature demonstrates a wide range of methodological skills.  In many papers, the analysis is confined to a narrow suite of familiar statistical methods such as least-squares regression (Legendre 1805), chi-squared test (Pearson 1901), Kolmogorov-Smirnov goodness-of-fit test (Kolmogorov, 1933), principal components analysis (Hotelling 1936).  Even these traditional methods can be misused.   For example, the astronomers' `minimum chi-squared' regression method has ill-determined degrees of freedom if it is based on arbitrarily binned data, according to an important  theorem (Chernoff \& Lehmann 1954).  The Kolmogorov-Smirnov statistic is less sensitive than the Anderson-Darling statistic that gives a tail-weighted Cramer-von Mises test (Stephens 1986).   Aperiodic autocorrelated behaviors in time series, such as long-memory $1/f^\alpha$ variability, is difficult to estimate in a reliable fashion.

The weak methodology in many studies can be directly attributable to the lack of formal training in statistics.  Astronomers usually take zero or one course in statistics during their university and graduate education, in contrast to extensive training in physics and associated mathematics.  Education is similarly weak in computational methods.  A survey of $\sim$1100 astronomers worldwide (Momcheva \& Tollerud 2015) shows that 90\% write software but only 8\% receive substantial training in software development and 43\% receive no training. 	Preferred languages are Python (67\%) and IDL (44\%), but only 6\% (3\% in USA) prefer R, the premier software environment for advanced statistical methods.  Momcheva and Tollerun remark that "considering the wide-spread use of R in other scientific fields, its popularity among astronomers is strikingly low?.  For this reason, the R language is briefly reviewed below (\S~4).  

The result of weak training is a widespread ignorance of both fundamental principles and of the myriad recent advances in modern statistical and computational methodology.  These fields are huge, so even if an astronomer is familiar with contemporary methods in one area, they may be unfamiliar with methodology in other areas.  

\section{The resurgence of astrostatistics and astroinformatics}

Fortunately, a growing fraction of the astronomical research community has recognized these deficiencies, and considerably activity in methodology has emerged in recent years.  Papers in the astronomical literature with the keywords "Methods: statistical" or "Methods: Numerical" have more than doubled in the past decade to $\sim$1000 studies annually, nearly 10\% of the total literature.    Nearly 2000 astronomers, many of them graduate students, have attended short (1-5 day) professional development tutorials in astrostatistics and astroinformatics in a dozen countries.  Cross-disciplinary research collaborations are active in several distinguished universities, particularly in the United States.  Cross-disciplinary conferences in astrostatistics and astroinformatics, ranging from topical workshops to sessions within large congresses to week-long symposia, have been held with increasing frequency.  

Several scholarly societies have taken note of the growing importance of advanced methodology for astronomy, particularly in light of the growth of Big Data surveys.  These include the International Astrostatistics Association affiliated with the International Statistical Institute, Commission on Astroinformatics and Astrostatistics within the International Astronomical Union, Interest Group in Astrostatistics within the American Statistical Association, Working Group in Astroinformatics and Astrostatistics within the American Astronomical Society, and the Astrominer Task Force within the Institute of Electrical and Electronics Engineers.  The Large Synoptic Survey Telescope, the U.S. national project, has a cross-disciplinary Informatics and Statistics Science Collaboration.  

New resources are available for students and researchers.  Textbooks that focus on statistical analysis of astronomical data are available with software implementations in Mathematics, Python and R (Gregory 2005, Wall \& Jenkins 2012, Feigelson \& Babu 2012, Ivezi\'c et al. 2014).  A large collection of recent papers, meetings, jobs, blogs and other on-line resources is available from the Astrostatistics and Astroinformatics Portal (asaip.psu.edu).  Up to forty gatherings worldwide occur annually related to these growing fields.  

\section{R: A powerful language for statistical analysis}

There is little practical utility for advanced statistical and computational methodology if software implementations are not available.  The astronomical community commits considerable resources to software specific to particular instruments and science goals, but these incorporate mostly familiar and simple methods.  The more sophisticated methods are often focused on particular science projects and are the responsibility of individual scientists, rather than the larger software teams committed to observatory data analysis.  It is in this later science analysis phase, illustrated in Table 1, rather than earlier pipelined data reduction phase that new software capabilities are needed.  Often the problems are not unique to a particular study, but embody analysis goals seen elsewhere in astronomy and other scientific fields.  Thus, general utility statistical packages can be quite useful for the science phase in astronomical studies.

In the era of mainframe computers, academic statisticians did not promulgate code but rather relegated the task to commercial software systems like SAS (Statistical Analysis System) and SPSS (Statistical Package for the Social Sciences).  During the 1980s when the C language was developed at AT\&T Labs, John Chambers at the Labs created the S statistical package written in C.  This became the commercial S-Plus package.  In the 1990s as the Internet emerged, New Zealand statisticians Ross Ihaka and Robert Gentleman rewrote the S language as an open source system called R.  The R Core Development Team of a dozen statistical computing experts around the world formed, and the system was released to the public under a GNU General Public License.  This included a procedure for users to contribute specialized packages within the Comprehensive R Analysis Network (CRAN).  

Unexpectedly, the contributed CRAN packages proliferated at an enormous rate, growing exponentially from 2001 through 2011.  Today 5-6 new CRAN packages are submitted every day, some with one function (subroutine) and others with hundreds.  Major fields with collections of disciplinary CRAN packages include genomics and econometrics but packages serve fields as diverse as actuarial science, archeology, clinical trials, earthquake, entomology, industrial quality control, Internet streams. Most packages implement advanced statistical methods without reference to particular disciplines.  In late-2016, R has about 10,000 CRAN packages with perhaps 150,000 functions including statistical operations, datasets and infrastructure.  About 20 packages are specifically oriented towards astronomy including FITS format input/output and a translation of much of the IDL Astronomy Library.   R and Python are the principal languages in the business fields of data science and analytics, and R dominates posts on software forums like stackoverflow and Talk Stats.  Thus, R has quickly grown to be the world's premier statistical computing environment that implements a large fraction of modern statistics.  It is larger than SAS, the most comprehensive of the commercial statistical packages, and is used by at least 2 million individuals.  

Astronomers have little difficulty learning the R language, as its syntax is quite similar to IDL and Matlab.  It is a compact scripting language where a statistical operation, which may be very complicated and require extensive computation, is called in a single line.  Interactive R provides a Graphical User Interface with several available integrated development environments. R has a byte code compiler similar to IDL, Python and Matlab.  All of these languages give rapid computation for vector/matrix inputs but are slower for more complex program structures like loops. Fortran, C or C++ code can be easily incorporated into R scripts for high performance computing.  Important to astronomy, two-way communication exists between R and Python; one can wrap an R/CRAN function in a Python code, or wrap a Python program within an R script.  

The main difficulty in using R is its size and diversity.  Astronomers need sufficient education in methodology to identify what procedure within R/CRAN is needed for a particular application.  The \textit{CRAN Task Views} give brief roadmaps ro R and CRAN capabilities in broad areas of statistics like Bayesian inference, machine learning or time series analysis.    For example, CRAN currently has 34 packages for wavelet analysis,  some disciplinary specific and others general toolboxes.  Some individual CRAN packages are very impressive.  The {\it spatstat} package gives over a thousand functions to analyze spatial point processes in 2 or 3 dimensions for datasets like galaxy redshift surveys.  Important packages are described in books such as the {\it spatstat} volume by Baddeley et al. (2015) or the dozens of cookbooks in Springer's \textit{Use R!} book series.  Dozens of other books give introductions to basic statistical analysis with R or, like Feigelson \& Babu (2012) and Hilbe et al. (2017) for astronomers, are guides for particular disciplines.  

While R was not respected for high performance computing in the past, dozens of recent CRAN packages link R to various modern computational hardware structures.  With the $foreach$ function, loops can be easily distributed among local cores. CRAN packages treat parallel supercomputers with Message Passing Interface, Parallel Virtual Machine, Open Multi-Processor, Apache Hadoop and other frameworks.  Other packages facilitate use of large datasets out of local memory (including commercial cloud computing), and GPU computing.

\section{The landscape: past, present and future}

Astronomy played a critical role in the foundation of modern statistics with the development of least squares regression in the early 19th century, just as it was at the foundation of modern physics with celestial mechanics of the 18-19th centuries.  But during the 20th century, astronomy and statistics parted ways.  The former developed astrophysics based on electromagnetism, quantum mechanics and other branches of physics, while the latter moved to serve human needs in industry, government, social and biological sciences.  By the late 20th century, the situation was poor: astronomers were unaware of important advances in methodology while statisticians were unaware of research issues in contemporary astronomy.  But the field of astrostatistics, both in learning established sophisticated methods and in pushing the envelope for new methodology inspired by astronomical research problems, exhibits strong growth during the 21st century.  

Astronomy did not play a central role in the development of computing machines and languages.  The innovations of Charles Babbage and Ada Lovelace, Alan Turing, Claude Shannon, John Von Neumann, Bill Gates, Ken Thompson and Dennis Ritchie (developers of Unix), and Richard Stallman (GNU pioneer) had little link to astronomy or other physical sciences.  Only David Stern, the developer of the Interactive Data Language since the 1970s, was motivated by astronomical data problems.  Rather, astronomers have become customers of computer hardware and software products and methodologies intended for business, government, engineering, and other purposes.  While the majority of astronomers use widely distributed standard hardware and software, a growing segment of the community has acquired skills in parallel processing, GPU computing, and machine learning.  Astroinformatics is a very new speciality that is also quickly emerging today.  

The landscape of these fields is thus changing rapidly.  The situation today feels somewhat disorganized as there is no natural center for these developments and they are widely dispersed geographically.  Although scholarly societies show strong interest in these new fields, funding agencies are only beginning to be responsive to their needs.  The pattern has been to fund `software' development as an engineering effort closely linked to astronomical instrumentation without much study of optimal statistical and computational `methods' underlying the software.  However, the external worlds of statistics, applied mathematics and computer science are so vast, and change so rapidly, that this informal approach can no confidently give rise to high quality procedures and outcomes.  Cross-disciplinary experts should be hired within, or should serve as outside consultants to, astronomical software teams.  Cross-disciplinary research groups in astrostatistics and astroinformatics should be nurtured at universities and institutes, just as astrochemistry and instrumentation groups with specialized skills emerged during the late 20th century.  

The impetus for quickly incorporating astrostatistics and astroinformatics into the astronomical research community is propelled by the emergence of superb new high-throughput telescopes culminating with the optical-infrared band Large Synoptic Survey Telescope and radio band Square Kilometre Array expected during the 2020s.  Advanced methodology has always been an opportunity for improving science analysis and results.  But in the Big Data era, it is essential for reaching the scientific potential of the new instruments.  

\medskip
The author is grateful for support by Penn State's Center for Astrostatistics and the National Science Foundation (AST-1614690).

\end{document}